\documentclass[12pt]{article}
\pdfoutput=1
\usepackage{latexsym, graphicx,cite,mathrsfs}
\usepackage{amsmath}
\usepackage{amssymb}
\usepackage{amsthm}
\usepackage{subcaption}

\allowdisplaybreaks

\usepackage[top = 1. in,bottom = 0.9 in, left = 1 in, right=1 in]{geometry}
\usepackage[colorlinks=true, linkcolor=blue, bookmarks=true]{hyperref}
\newcommand{\arXiv}[1]{\href{http://www.arXiv.org/abs/#1}{arXiv:#1}}
\newcommand{\doi}[2]{\href{http://dx.doi.org/#2}{#1}}

\makeatletter
\renewcommand\section{\@startsection {section}{1}{\z@}%
                  {-3.5ex \@plus -1ex \@minus -.2ex}
                  {2.3ex \@plus.2ex}%
                  {\normalfont\large\bfseries}}
\renewcommand\subsection{\@startsection{subsection}{2}{\z@}%
                   {-3.25ex\@plus -1ex \@minus -.2ex}%
                   {1.5ex \@plus .2ex}%
                   {\normalfont\bfseries}}
\makeatother

\newcommand{\beq}{\begin{equation}}
\newcommand{\eeq}{\end{equation}}
\newcommand{\ber}{\begin{array}}
\newcommand{\eer}{\end{array}}
\newcommand{\del}{\partial}

\newcommand{\ph}{\varphi}
\newcommand{\de}{\delta}
\newcommand{\la}{\lambda}

\newcommand{\bea}{\begin{eqnarray}}
\newcommand{\eea}{\end{eqnarray}}
\newcommand{\ER}{Erd\H{o}s-R\'enyi }
\renewcommand{\Re}{\operatorname{Re}}
\renewcommand{\Im}{\operatorname{Im}}
\newcommand{\Tr}{\mathrm{Tr}}
\newcommand{\Prj}{\mathbb{P}}

\newcommand{\mbf}{\mathbf}
\newcommand{\mbb}{\mathbf}
\newcommand{\dagg}{\dagger}

\newcommand{\be}{\beta}
\newcommand{\ga}{\gamma}
\newcommand{\nn}{\nonumber}

\begin{document}
\begin{titlepage}
\begin{flushright}
\phantom{arXiv:yymm.nnnn}
\end{flushright}
\vspace{5mm}
\begin{center}
{\LARGE\bf\mbox{\rule{-7mm}{0mm}Hammerstein~equations~for~sparse~random~matrices}}\\
\vskip 12mm
{\large Pawat Akara-pipattana$^{a}$ and Oleg Evnin$^{b,c}$}
\vskip 10mm
{\em $^a$ Universit\'e Paris-Saclay, CNRS, LPTMS, 91405 Orsay, France}
\vskip 3mm
{\em $^b$ High Energy Physics Research Unit, Department of Physics, Faculty of Science, Chulalongkorn University,
10330 Bangkok, Thailand}
\vskip 3mm
{\em $^c$ Theoretische Natuurkunde, Vrije Universiteit Brussel and\\
 International Solvay Institutes, 1050 Brussels, Belgium}
\vskip 7mm
{\small\noindent {\tt pawat.akarapipattana@universite-paris-saclay.fr, oleg.evnin@gmail.com}}
\vskip 20mm
\end{center}
\begin{center}
{\bf ABSTRACT}\vspace{3mm}
\end{center}

Finding eigenvalue distributions for a number of sparse random matrix ensembles can be reduced to solving nonlinear integral equations of the Hammerstein type.
While a systematic mathematical theory of such equations exists, it has not been previously applied to sparse matrix problems. We close this gap in the literature by showing how one can employ numerical solutions of Hammerstein
equations to accurately recover the spectra of adjacency matrices and Laplacians of random graphs. While our treatment focuses on random graphs for concreteness, the methodology has broad applications to more general sparse random matrices.

\vfill

\end{titlepage}

\section{Introduction}

The spectra of random matrices have found far-reaching applications in situations where large complex matrices are involved, including fields as diverse as 
linear \cite{May,modu} and nonlinear \cite{nonlin} dynamics in the presence of random couplings, quantum chaos \cite{chaos}, localization phenomena in disordered media \cite{anderson}, vibrational properties of amorphous solids \cite{amorph}, the physics of heavy nuclei \cite{nuclei},
explorations of discrete geometry \cite{discr} and mathematical neurobiology \cite{neuro}. One particular topic is spectral problems related to graphs, 
and hence to network theory \cite{newman},
since graphs are connected to matrices via the adjacency matrix representation; see \cite{VM} for a textbook exposition.

A number of spectral problems for {\it sparse} random matrices, that is, $N\times N$ matrices with a large $N$ and a fixed average number $c$ of nonzero entries per row,
can be reduced \cite{BR,RB,FM,MF,gel,Khetal,inteq,laplace,inhmg,walkongraph} to solving nonlinear integral equations of the form
\beq\label{Hammdef}
g(x)=s(x)+\int dx'\, K(x,x') F(x',g(x')),
\eeq
where the `source' $s(x)$, the kernel $K(x,x')$ and the nonlinearity $F$ are prescribed, the latter typically being an exponential nonlinearity with respect to the unknown $g(x)$ in the examples we will consider. Such equations are known as Hammerstein equations, after \cite{hmmr}.  A typical reaction of a theoretical physicist to an equation like (\ref{Hammdef}) is that of bewilderment. We thus read in the influential paper \cite{kuehn} on sparse matrix spectra: ``This integral equation...\ has, however, so far resisted exhaustive analysis or full numerical solution.'' An early numerical treatment can be seen in \cite{gel}, and it approximates the desired solution via sophisticated sums of Gaussians adapted to the concrete equation at hand, without evoking a systematic theory of equations of the form (\ref{Hammdef}).

Despite having remained unincorporated in the context of sparse random matrix considerations, a systematic theory of Hammerstein equations has been developed in the mathematical literature, starting with \cite{hmmr}, see \cite{surv} for a comprehensive review. (Curiously, the physical application
of such equations mentioned in \cite{surv} is in Chandrasekhar's theory of radiative transfer \cite{Chandra}, very far from our present perspective.) In fact, a splash of research activity on numerical convergence
properties in concrete cases is seen in the recent years. Our purpose in this article is to demonstrate that this systematic approach to solving Hammerstein
equations numerically provides an efficient procedure for recovering sparse matrix spectra in a way that has been previously overlooked in the physics literature.
The concrete examples we shall consider are the spectra of adjacency matrices, ordinary graph Laplacians and normalized graph Laplacians of sparse \ER random graphs. These examples 
are chosen as concrete random matrix ensembles that have received considerable attention in the literature. The technology we discuss should be applicable with minimal modifications to many other sparse random matrices.

The paper is organized as follows: In section~\ref{seqSFT}, we review the emergence of integral equations from the problem of computing the spectrum
of adjacency matrices of sparse \ER graphs. In section~\ref{seqHamm}, we describe the process for solving Hammerstein equations numerically. In section~\ref{seqnum}, we present various tweaks necessary for handling the concrete Hammerstein equations arising from random matrices, as well as the practical results
of our numerical work for the spectra of adjacency matrices, ordinary graph Laplacians and normalized graph Laplacians of sparse \ER random graphs.
We conclude with a discussion and a wish\-list for possible improvements in the numerical algorithms.

\section{An integral equation for sparse random graph spectra}\label{seqSFT}

We start with a review of the emergence of integral equations of Hammerstein type from the application of statistical field theory methods \cite{statFT,statFTneu} to spectral problems for sparse random graphs. These methods have been used for a wide range
of problems involving random graphs and random matrices \cite{RB,FM,MF,kuehn,BR,euclRM,SC,PN,spectrum,AC,TO,RP2022,resdist,Baron}. Our purpose here is not to provide a systematic technical exposition of these methods, but rather to give the readers an impression of the sort of structures and steps involved.
The readers have the option of fast-forwarding through this section, as the topic of deriving the Hammerstein equations for sparse random matrix spectra is
logically independent from the topic of solving them.

The specific variation of the statistical field theory methods we use relies on supersymmetry and auxiliary integrals over anticommuting variables,
see \cite{efetov, wegner} for textbook expositions. In application to random matrices, some essential ingredients we use go back to Fyodorov and Mirlin \cite{FM,MF}. We have given a pedagogical account of this technology in \cite{laplace}, and refer the readers to that paper for further details. 

The example we choose for our demonstration is the eigenvalue density of the adjacency matrix of an \ER random graph. This graph is defined as a set of $N$ vertices with $N\gg 1$, and each pair of vertices is connected randomly and independently with probability $c/N$, so that at $N\to\infty$, each vertex has on average $c$ connections. These data are conveniently coded into the {\it adjacency matrix} $\mathbf{A}$, which is a symmetric zero-diagonal matrix whose entry $A_{ij}$ equals 1 if vertices $i$ and $j$ are connected, and 0 otherwise. The above specification of the \ER ensemble is translated into the following probability distribution for the entries of $\mathbf{A}$:
\beq\label{adjdef}
P(\mathbf{A})=\prod_i \de(A_{ii})\,\,\prod_{i<j}\left[\left(1-\frac{c}{N}\right)\de(A_{ij})+ \frac{c}{N}\de(A_{ij}-1)\right]\de(A_{ji}-A_{ij}),
\eeq
where $\de$ stands for the standard $\de$-function. Finding the eigenvalue density of the adjacency matrices described by this ensemble is a standard problem
in random matrix theory. Our treatment follows the original works \cite{MF,FM}. We choose this example for its relative simplicity, and also for the fact that it is not directly covered by the pedagogical exposition of \cite{laplace}. At large $c$, the eigenvalue distribution tends to the semicircle familiar from many incarnations of random matrix theory \cite{RMTrev}. At smaller $c$, the shape gets deformed in a complicated way, and it is this deformed shape that will
be captured by integral equations we consider in this paper.

We start with the so-called {\it resolvent} representation of the eigenvalue density $p(\la)$. Let the eigenvalues of $\mathbf{A}$ be $\la_i$. 
Then, by definition:
\beq\label{rhodef}
p(\la)\equiv\frac1N\sum_{k=1}^N\de(\la-\la_k)=\frac1{\pi N}\Im\sum_k\frac1{\la-\la_k-i0}=-\frac1{\pi N}\Im\Tr[\mathbf{A}-z\mathbf{I}]^{-1}\Big|_{z=\la-i0},
\eeq
where we have used the Sokhotski–Plemelj formula and the spectral decomposition of the matrix inverse, while $z=\la-i0$ implies evaluating the relevant expression just below the real axis. The resolvent $\Tr[\mathbf{A}-z\mathbf{I}]^{-1}$ is much simpler to handle within statistical averages than the eigenvalues themselves, since it can be effectively represented using Gaussian integrals, making the dependence on the components of $\mathbf{A}$ simplify further.
Note first that, due to the symmetry of the \ER ensemble under vertex permutations, after the averaging, one has
\beq\label{ressimpl}
\frac1N \langle\Tr[\mathbf{A}-z\mathbf{I}]^{-1}\rangle\equiv\frac1N \sum_{k=1}^N \langle[\mathbf{A}-z\mathbf{I}]^{-1}_{kk}\rangle= \langle[\mathbf{A}-z\mathbf{I}]^{-1}_{11}\rangle.
\eeq

The next step is to represent the resolvent (\ref{ressimpl}) through Gaussian integrals, which results in factorization over the entries of $\mathbf{A}$ and hence in a straightforward $\mathbf{A}$-averaging. To do so, it is customary \cite{MF,FM,efetov,wegner} to introduce a 4-component `supervector' at every vertex $k$, given by $\Psi_k\equiv (\phi_k,\chi_k,\xi_k,\eta_k)$, where $\phi_k$ and $\chi_k$ are ordinary numbers, while $\xi_k$ and $\eta_k$ are anticommuting numbers satisfying
\beq\label{Berezin}
\xi_i\xi_j=-\xi_j\xi_i,\qquad \xi_i^2=0,\qquad \int d\xi_i \xi_j=\de_{ij},\qquad \int d\xi_i=0,
\eeq
the latter relations known as Berezin integration rules. (All components of $\xi$ anticommute with all components of $\eta$ as well. Note that Berezin integrals should not be visualized as any generalization of Riemann sums. They are linear maps from functions of anticommuting numbers to real numbers that have been developed historically for the purpose of convenient path-integral representation of fermionic systems.)
We furthermore introduce a `scalar product' of supervectors defined as
\beq\label{scalarp}
\Psi^\dagg_k\Psi_l\equiv \phi_k\phi_l + \chi_k\chi_l +\frac{\xi_k\eta_l-\eta_k\xi_l}{2},\qquad (\Psi^\dagg_k\Psi_k\equiv \phi_k^2 + \chi_k^2 +\xi_k\eta_k).
\eeq
With these preliminaries, we can write
\begin{equation}\label{Gauss}
(\mbf{A} - z\mbb{I})^{-1}_{11} = \int d\mbf{\Psi}\, (2i\phi_1^2) \,e^{i\sum_{kl}A_{kl}\Psi^\dagg_k\Psi_l - iz\sum_k \Psi^\dagg_k\Psi_k}.
\end{equation}
Note that, if substituting (\ref{scalarp}) into (\ref{Gauss}), one obtains independent Gaussian integrals involving $\exp(i\sum A_{kl}\phi_k\phi_l-iz\sum \phi_k^2)$,  $\exp(i\sum A_{kl}\chi_k\chi_l-iz\sum \chi_k^2)$ and $\exp(i\sum A_{kl}\xi_k\eta_l-iz\sum \xi_k\eta_k)$. Absorbing the powers of $\pi$ arising from the standard Gaussian integrals over $\phi$ and $\chi$ into the definition of the measure, the first integral evaluates to $(\mbf{A} - z\mbb{I})^{-1}_{11}/\sqrt{\det(\mbf{A} - z\mbb{I})}$ due to the insertion of $\phi_1^2$, the second integral evaluates to $1/\sqrt{\det(\mbf{A} - z\mbb{I})}$, while the last integral evaluates to $\det(\mbf{A} - z\mbb{I})$ by the standard formula for Gaussian integrals in Berezin integration theory. As a result, the powers of determinants exactly cancel, leaving behind (\ref{Gauss}).

The purpose of inserting integrals over anticommuting variables in (\ref{Gauss}) was precisely to make the powers of determinant arising from Gaussian integration exactly cancel out, yielding an expression factorized over the components of $\mathbf{A}$. An alternative strategy, often seen in statistical physics literature, is the replica method \cite{RB,BR, kuehn,RP2022,suscaetal}, where one introduces an {\it arbitrary} controllable number of extra Gaussian integrals
over ordinary commuting variables, formally extrapolating the total number of Gaussian integrals to 0 in the final formulas at the end of the computation. While, in practical terms, this strategy is largely interchangeable with the supersymmetry-based approach we are employing here, the latter dispenses of the need for the ambiguous operation
of extrapolating a positive integer to 0.

Averaging (\ref{Gauss}) over the random matrix ensemble (\ref{adjdef}) is straightforward and yields
\begin{align}
\left<(\mbf{A} - z\mbb{I})^{-1}_{11}\right>& =\int d\mbf{\Psi}\, (2i\phi_1^2) \,e^{ - iz\sum_k \Psi^\dagg_k\Psi_k} \prod_{k<l}\left[\left(1-\frac{c}{N}\right)+\frac{c}{N}e^{2i\Psi^\dagg_k\Psi_l}\right]\nn\\
&=\int d\mbf{\Psi}\, (2i\phi_1^2) \,e^{-\frac{1}{2N}\sum_{kl}C(\Psi_k,\Psi_l) - iz\sum_k \Psi^\dagg_k\Psi_k},\label{superv}
\end{align}
where we have used $c/N\ll 1$ and then introduced 
\beq
C(\Psi,\Psi') = c\left(1-e^{2i\Psi^\dagg_k\Psi_l}\right).
\eeq
To solve the supervector model (\ref{superv}) at large $N$, we employ, after Fyodorov and Mirlin \cite{FM,MF,laplace}, the following Gaussian functional integral over an auxiliary function $g(\Psi)$:
\beq\label{MFtransfrm}
\int \mathcal{D}g \exp\left[-\frac{N}2\int d\Psi\,d\Psi'\, g(\Psi) \,C^{-1}(\Psi,\Psi')\,g(\Psi')+i\sum_k g(\Psi_k)\right]=e^{-\frac1{2N}\sum_{kl}C( \Psi_k, \Psi_l)},
\eeq
where $C^{-1}$ is the inverse of $C$ in the sense of integral convolution: 
\beq
\int d\Psi\, C(\Psi_1,\Psi)\,C^{-1}(\Psi,\Psi_2)=\de(\Psi_1-\Psi_2).
\eeq
The explicit form of $C^{-1}$ will not be necessary for our derivation. Once this transformation has been implemented, the integrals over $\Psi_k$ with different $k$ completely factorize and are performed independently, yielding an explicit saddle point problem:
\beq
\label{afterMF}
\begin{split}
&\langle(\mathbf{A}-z\mathbf{I})^{-1}_{11}\rangle= \int \mathcal{D}g\,e^{-N\hspace{0.1mm}S[g]}\,\frac {\int d\Psi\,(2i\phi^2)\,e^{ig(\Psi)-iz\Psi^\dagger\Psi}}{\int d\Psi\,e^{ig(\Psi)-iz\Psi^\dagger\Psi}},\\
&S[g]\equiv\frac{1}2\int d\Psi\,d\Psi'\, g(\Psi) \, C^{-1}(\Psi,\Psi')\,g(\Psi')-\ln\left(\int d\Psi\,e^{ig(\Psi)-iz\Psi^\dagger\Psi}\right).
\end{split}
\eeq
At large $N$, the integral is dominated by the stationary points of the functional $S[g]$ defined by $\de S/\de g(\Psi)=0$, which is 
\beq\label{sddlpsi}
g(\Psi)=i\,\frac{\int d\Psi'\,C(\Psi,\Psi')\,e^{ig(\Psi')-iz\Psi'^\dagger\Psi'}}{\int d\Psi'\,e^{ig(\Psi')-iz\Psi'^\dagger\Psi'}}.
\eeq
This equation is closely reminiscent of the analogous Bray-Rodgers equation in the replica approach \cite{BR,RB,kuehn}.

The final step is to assume that the dominant saddle point described by (\ref{sddlpsi}) is invariant under `superrotations' that preserve the scalar product $\Psi^\dagg\Psi$ (an alternative would have been continuous families of saddle points related by superrotations, since the equation is symmetric under applying a superrotation to the argument of $g$). This implies that $g(\Psi)=g(\Psi^\dagg \Psi)$, and with this simplification, the integral equation (\ref{sddlpsi}) can be reduced to a single one-dimensional integral equation. To do so, we express the commuting components of $\Psi$ in polar coordinates as
\beq\label{polar}
\rho=\phi^2+\chi^2,\qquad \phi=\sqrt{\rho}\cos\alpha,\qquad \chi=\sqrt{\rho}\sin\alpha,\qquad d\phi\,d\chi=\frac12 d\rho\, d\alpha.
\eeq
With this,
\beq\label{supersymm}
\Psi^\dagg\Psi=\rho+\xi\eta,\qquad g(\Psi^\dagg\Psi)=g(\rho)+\del_\rho g(\rho)\,\xi\eta,
\eeq
where the Taylor expansion of $g(\rho+\xi\eta)$ truncates at the $\xi\eta$-term, since all higher powers of anticommuting variables vanish by (\ref{Berezin}).
Inserting this representation in the $\Psi'$-integral in (\ref{sddlpsi}) and performing the integration over all variables except for $\rho'$, we obtain, after some integration by parts in $\rho'$:
\begin{equation}\label{sddlrho}
g(\rho) = ic\rho\int_0^\infty \!\!\!d\rho'\,\frac{J_1(2\sqrt{\rho\rho'})}{\sqrt{\rho\rho'}}\,e^{ig(\rho')-iz\rho'},
\end{equation}
where $J_1$ is the usual Bessel function.
This equation, which dates back to \cite{RB} where it was derived by replica methods, and its analogs for graph Laplacians, will be the main subjects of our numerical work in what follows. Similar integral equations with exponential nonlinearities and Bessel kernels arise also in other sparse random matrix problems.

Once a solution of (\ref{sddlrho}) has been found, one has to substitute it in (\ref{afterMF}) and perform a saddle point evaluation of the integral.
This, in general, would involve complicated determinants from Gaussian integrations, but in our case, these determinants must cancel out. An elementary way
to see it is as follows: consider, tautologically, (\ref{superv}) with $(\mbf{A} - z\mbb{I})^{-1}_{11}$ replaced by 1 (so that the average is guaranteed to equal 1).
Then, our entire derivation will go through, arriving at (\ref{afterMF}), but without the insertion $ {\int d\Psi\,(2i\phi^2)\,e^{\{\cdots\hspace{-0.2mm}\}}}/{\int d\Psi\,e^{\{\cdots\hspace{-0.2mm}\}}}$. Neither the saddle point equation (\ref{sddlrho}) nor the determinants arising from Gaussian integrations depend on the presence of this insertion. On the other hand, by our tautological starting point, the result without the insertion must equal 1. The result with the insertion, as in (\ref{afterMF}), must then be equal to the insertion evaluated at the saddle point, that is,
\beq
\langle(\mathbf{A}-z\mathbf{I})^{-1}_{11}\rangle=\frac {\int d\Psi\,(2i\phi^2)\,e^{ig(\Psi^\dagger\Psi)-iz\Psi^\dagger\Psi}}{\int d\Psi\,e^{ig(\Psi^\dagger\Psi)-iz\Psi^\dagger\Psi}},
\eeq
where $g$ is now understood as the solution of (\ref{sddlrho}). The integrals can again be simplified using (\ref{polar}-\ref{supersymm}) and assuming that $g(\rho)$ decays at infinity, and then plugged back into (\ref{rhodef}-\ref{ressimpl}) to yield the following result for the eigenvalue density:
\beq\label{extractp}
p(\lambda)=\frac1\pi\,\Re\int_0^\infty \!\!\!d\rho\, e^{ig(\rho)-iz\rho}\Big|_{z=\la}.
\eeq
By construction, $p(\lambda)$ is normalized as a probability density $\int d\la \,p(\la)=1$. It is rather striking, at the face value, that the solution of the
formidable equation (\ref{sddlrho}) processed with (\ref{extractp}) is automatically normalized in this way. It is nonetheless correct, as follows from our arguments, and as will be verified by numerical simulations. (In \cite{laplace}, we ignored all normalization at the intermediate stages of computation, since 
it is always easy to normalize the end result. Here, however, we restore all normalizations for comparisons with numerics, which can be done, as it turns out, with minimal effort.)


\section{Handling Hammerstein equations}\label{seqHamm}

Equation (\ref{sddlrho}) evidently matches the general structure of Hammerstein equations given by (\ref{Hammdef}).
Before proceeding with its solution, we will review the general underlying principles, following \cite{surv}.

In the random matrix literature, the systematic theory of Hammerstein equations appears to have never been applied.
Indeed, in \cite{kuehn}, the apparent intractability of integral equations like (\ref{sddlrho}) is used to motivate an alternative approach,
where one goes back to an analog of (\ref{sddlpsi}) and then develops a sampling scheme in the spirit of `population dynamics' to
approximate its solutions. In \cite{gel}, equations analogous to (\ref{sddlrho}) are approached directly, and their solutions are approximated
by elaborate sums of Gaussians. This approach is not apparently connected, however, to the strategies for solving Hammerstein equations
seen in the mathematical literature.

What the mathematical literature instructs us to do is to take (\ref{Hammdef}) and approximate its solutions by functions within some finite-dimensional
linear subspace spanned by $\phi_j(x)$ with $j=1..J$, that is $g(x)=\sum_j \beta_j \phi_j(x)$. One can then replace the original equation (\ref{Hammdef}) by the following modification
\beq\label{Hammtrunc}
\sum_j \beta_j \phi_j(x)=\Prj \left[s(x)+\int dx'\, K(x,x') F(x',{\textstyle\sum_j} \beta_j \phi_j(x'))\right],
\eeq
where $\Prj $ is a projector on the subspace spanned by $\phi_j(x)$. There are different possible choices for this projector, leading to different numerical solution methods, as we shall discuss shortly. One can view (\ref{Hammtrunc}) as a system of $J$ nonlinear algebraic equations for $J$ unknowns $\beta_j$ that can be solved by any standard numerical methods. One expects that, as the dimension $J$ of the subspace spanned by $\phi_j(x)$ increases,
the resulting $g(x)=\sum_j \beta_j \phi_j(x)$ will approximate the true solution of (\ref{Hammdef}) better and better. The details of this convergence have been studied by mathematicians, but we shall be pragmatic about it since rather small values of $J$ around 10 will be sufficient to reproduce sparse matrix spectra with a good precision.

There is a technical issue with equation (\ref{Hammtrunc}) as it stands. Because $\beta_j$ appear inside the integral over $x'$ nonlinearly, the integral
will have to be recomputed for each assignment of $\beta_j$, for example, when solving the algebraic system (\ref{Hammtrunc}) by means of iterative methods.
These repeated integrations are numerically costly. Kumar and Sloan proposed in \cite{KumarSloan} a simple trick to bypass this issue. Instead of $g(x)$,
one introduces 
\beq
f(x) \equiv F(x,g(x)),
\eeq
so that (\ref{Hammdef}) is equivalently rewritten as
\begin{equation}\label{KS}
f(x) = F\left(x,s(x) + \int dx'\, K(x,x')f(x')\right).
\end{equation}
We then expand $f(x)$, rather than $g(x)$, in the truncated basis $\phi_j(x)$ with $j=1..J$:
\beq
f(x)=\sum_{j=1}^J  \beta_j \phi_j(x),
\eeq 
and project (\ref{KS}), rather than (\ref{Hammdef}), on this basis using a suitable projector $\Prj $:
\begin{equation}\label{KStrunc}
\sum_j \beta_j \phi_j(x)=\Prj F\left(x,s(x) +{\textstyle \sum_j} \beta_j \int dx'\, K(x,x') \phi_j(x')\right).
\end{equation}
This is, again, solved for $\beta_j$ as a system of nonlinear algebraic equations. An advantage over (\ref{Hammtrunc}) is that the integrals
$\int dx'\, K(x,x') \phi_j(x')$ are computed once and for all, and are not affected by the values of $\beta_j$.
After $\beta_j$ has been found in this way, providing an approximation to $f(x)$, $g$ is best reconstructed by applying one more integration:
\begin{equation}
g(x) =  s(x) + \int dx' \,K(x,x')f(x').
\end{equation}

Regarding the choice of the projector $\Prj $, a natural first thought is to use an orthogonal projector with a suitably defined inner product
on the space of functions. This choice leads to Galerkin (or spectral) methods for solving Hammerstein equations. These methods are extensively covered in the literature, but a more economical setup in the context of (\ref{KStrunc}) is given by the {\it collocation} method that uses the so-called interpolating projector
\cite{Legendre}. For a given input $h(x)$, the interpolating projector provides $\Prj h(x)$ as a function of the form $\sum_{j}  \beta_j \phi_j(x)$ that equals $h(x)$ at prescribed collocation points $x_k$ with $k=1..J$. This provides exactly $J$ linear conditions for $J$ unknowns $\beta_j$. With such a prescription for the projector $\Prj $, (\ref{KS}) becomes
\begin{equation}\label{colloc}
\sum_j \beta_j \phi_j(x_k)= F\left(x_k,s(x_k) +{\textstyle \sum_j} C_{kj}\beta_j \right),\qquad C_{kj}\equiv\int dx\, K(x_k,x) \phi_j(x).
\end{equation}
Note that the $J\times J$ matrix $C_{kj}$ only has to be evaluated once. With this matrix computed, the $J$ equations for $\beta_j$, indexed by $k$, are solved using the standard numerical routines for multidimensional root search.

One still has to decide on how to choose the basis $\phi_j$ and the collocation points $x_k$. Mathematical literature consistently advises us to work
with orthogonal polynomials. In \cite{Legendre}, integral equations on the interval $[-1,1]$ are considered, and then it is natural to choose $\phi_j$ as the Legendre polynomials. Some considerations of half-infinite integration ranges, as in (\ref{sddlrho}), can be seen in \cite{Laguerre}. Since the integrals are from $0$ to $\infty$, it is natural to use Laguerre
polynomials but we will make some different choices in the concrete examples below, depending on the concrete form of the integrand appearing in the Hammerstein equation. Note that since no orthogonal projection is involved in the collocation equation (\ref{colloc}), the orthogonality property of the polynomial family is never used explicitly, and any other family of linearly independent functions could be used, at least in principle. Using orthogonal polynomials is, however, a way to ascertain a `good degree' of linear independence. For example we have checked that constructing $\phi_j$ from plain polynomials $1$, $x$, $x^2$, etc without orthogonalization results in a poorly conditioned system, and the numerical search for the solutions of (\ref{colloc}) does not reliably converge.
Orthogonal polynomials work well, by contrast. In \cite{Laguerre}, we are instructed that a good choice of collocation points is the roots of the lowest-degree orthogonal polynomial not included in the chosen set $\phi_j$. We will generally follow this prescription, though according to our
simulations reported below, reasonable variations in the positions of the collocation points do not immediately compromise the operation of the algorithm.


\section{Numerical results for sparse \ER graphs}\label{seqnum}

We proceed to report how numerical solutions of Hammerstein equations via the collocation method (\ref{colloc}) work for determining
eigenvalue spectra of adjacency matrices and Laplacians of sparse random graphs. For each concrete equation at hand, we will have to indicate
how to effectively partition the functional dependences in the integral convolutions on the right-hand side to match the notation in (\ref{Hammdef}), and how to choose the sets of trial functions
$\phi_j$ and collocation points.

\subsection{Adjacency matrices}\label{subseqadj}

Before applying the technology of section~\ref{seqHamm} to equation (\ref{sddlrho}), it is wise to implement the following rescalings:
\begin{equation}\label{scaleadj}
\rho = \tilde{\rho}/\sqrt{c}, \hspace{0.5cm} z = \tilde z\sqrt{c}, \hspace{0.5cm} g(\rho(\tilde\rho)) =  \tilde{g}(\tilde\rho).
\end{equation}
Dropping the tildes for convenience, we obtain the rescaled equation
\begin{equation}\label{adjrescaled}
g(\rho) = i\rho\int_0^\infty \hspace{-2mm} d\rho'\, \frac{J_1(2\sqrt{\rho\rho'/c})}{\sqrt{\rho\rho'/c\,}} \,e^{ig(\rho')-iz\rho'} .
\end{equation}
An advantage of this representation is that the relevant range of $\rho$ becomes approximately independent of $c$. Indeed, at large $c$, we can expand $J_1(2x)=x+\cdots$ and the equation turns into $g(\rho) = i\rho\int d\rho' \,e^{ig(\rho')-iz\rho'}$, solved by a fixed-size Wigner semicircle \cite{laplace}. 
At smaller $c$, the functional dependences evidently get deformed, but they occupy roughly the same region with respect to the new $\rho$ variable, rather than getting scaled $\sim\sqrt{c}$ with respect to the old $\rho$ variable. This makes the numerical implementation more neat.

There is some ambiguity in how to split the various $\rho$-dependences in (\ref{adjrescaled}) to match the structure of (\ref{Hammdef}).
Evidently, $s(\rho)=0$. We find it convenient to identify
\beq
K(\rho,\rho')=i\rho\,\frac{J_1(2\sqrt{\rho\rho'/c})}{\sqrt{\rho\rho'/c\,}} ,\qquad F(\rho,g(\rho))=e^{ig(\rho)-iz\rho}. 
\eeq
An advantage of including the $z$-dependence into the nonlinearity $F$ and not into the kernel $K$ is that, as we vary $z$ to derive the eigenvalue density at different points, the integrals $C_{kj}$ in the Kumar-Sloan collocation problem (\ref{colloc}) will not have to be recomputed. For the basis $\phi_j$ we choose Laguerre polynomials multiplied by exponentials. Namely, we write
\beq\label{expexp}
e^{ig(\rho)-iz\rho}=\sum_{j=0}^J \beta_j L_j(\ga\rho) \,e^{-\ga\rho},
\eeq
where $L_j$ are Laguerre polynomials orthogonal with respect to $\int_0^\infty dx\, e^{-x}L_i(x)L_j(x)$. We will treat $\ga$ as a tunable real parameters and adjust it to optimize the numerical performance. We remark that we could have used different scalings for the arguments of the Laguerre polynomials and the exponential, which gives more control over the numerical performance, but the practical gain is not significant and the formulas become more bulky. We give a summary of that scheme in Appendix~\ref{appexp}, and proceed here with (\ref{expexp}) as given. 

We can evaluate the integral in (\ref{adjrescaled}) analytically using the identity \cite{Sz}
\beq\label{Besselid}
\sum_{n=0}^\infty \frac{w^n\,L_n^{(k)}(x)}{(n+k)!}=e^w(xw)^{-k/2}J_k(2\sqrt{xw}),
\eeq
where $L^{(k)}$ are the associated Laguerre polynomials,
and choosing $x=\gamma\rho'$, $w=\rho/\gamma c$ and $k=1$. From (\ref{adjrescaled}), (\ref{expexp}) and (\ref{Besselid}), together with  $L^{(1)}_n=\sum_{m=0}^n L_m$,
we obtain 
\begin{align}
g(\rho)&=i\rho\, e^{-\rho/\ga c}\sum_{n=0}^\infty\frac{(\rho/\ga c)^n}{\ga (n+1)!}\sum_{m=0}^{\min(n,J)}\be_m\nn\\
&=i\rho\, e^{-\rho/\ga c}\left[\sum_{m=0}^{J}\be_m\sum_{n=0}^\infty\frac{(\rho/\ga c)^n}{\ga (n+1)!}-\sum_{n=0}^{J-1}\frac{(\rho/\ga c)^n}{\ga (n+1)!}\sum_{m=n+1}^{J}\be_m\right]\nn\\
&=ic\left[(1-e^{-\rho/\ga c})\sum_{m=0}^{J}\be_m- e^{-\rho/\ga c}\sum_{n=1}^{J}\frac{\rho^n}{ n!(\ga c)^n}\sum_{m=n}^{J}\be_m\right]\nn\\
&=ic\sum_{m=0}^{J}\be_m\left(1- e^{-\rho/\ga c}\sum_{n=0}^{m}\frac{(\rho/\ga c)^n}{ n!}\right).\label{grec}
\end{align}
Note that the $\rho'$-integral is completely gone due to the orthogonality of Laguerre polynomials.
Recombined with (\ref{expexp}), this generally results in the
following problem:
\beq\label{projexp}
\sum_{j=0}^J \beta_j L_j(x) =\Prj \exp\left[x-\frac{izx}{\ga}-c\sum_{j=0}^{J}\be_j\left(1- e^{-x/\ga^2 c}\sum_{n=0}^{j}\frac{(x/\ga^2 c)^n}{ n!}\right)\right],
\eeq 
where we defined $x\equiv \ga\rho$, and $\Prj $ is the projector of choice on the space of degree $J$ polynomials in terms of $x$.
This should be treated as a system of $J+1$ equations, from identifying the polynomial coefficients on the two sides, for $J+1$ unknowns $\be_j$. By orthogonality of Laguerre polynomials, (\ref{extractp}) then yields for the following eigenvalue density, keeping in mind that we must undo the rescalings (\ref{scaleadj}):
\beq\label{padjbeta}
p_\mathrm{adj}(\la)=\frac{\Re \be_0\big|_{z=\la/\sqrt{c}}}{\gamma\pi\sqrt{c}}.
\eeq
For the simplest choice of $\Prj $ we take the interpolating projector, equating the polynomial and the exponential in (\ref{projexp}) at the collocation points $x_k$ with $k=0..J$, chosen as the roots of $L_{J+1}$:
\beq\label{collocadj}
\sum_{j=0}^J \beta_j L_j(x_k) =\exp\left[x_k-\frac{izx_k}{\ga}-c\sum_{j=0}^{J}\be_j\left(1- e^{-x_k/\ga^2 c}\sum_{n=0}^{j}\frac{(x_k/\ga^2 c)^n}{ n!}\right)\right].
\eeq 
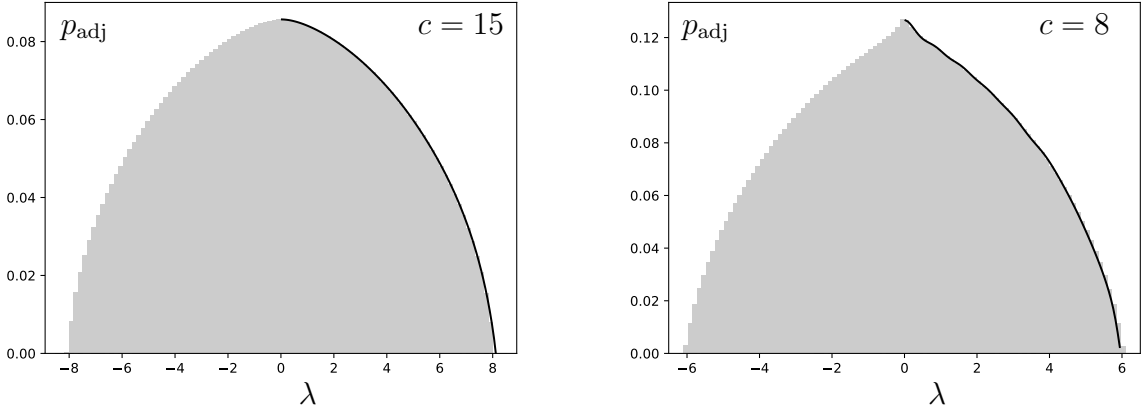
\begin{figure}[t]\vspace{-10mm}
\centering
\includegraphics[width = 0.49\linewidth]{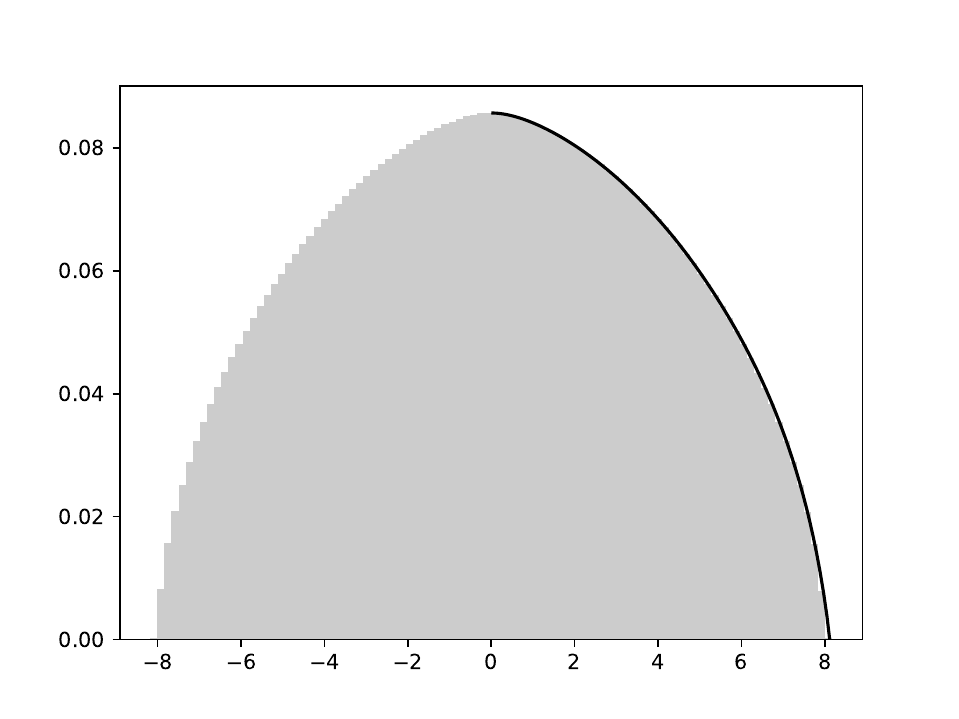}\hspace{2mm}\includegraphics[width = 0.49\linewidth]{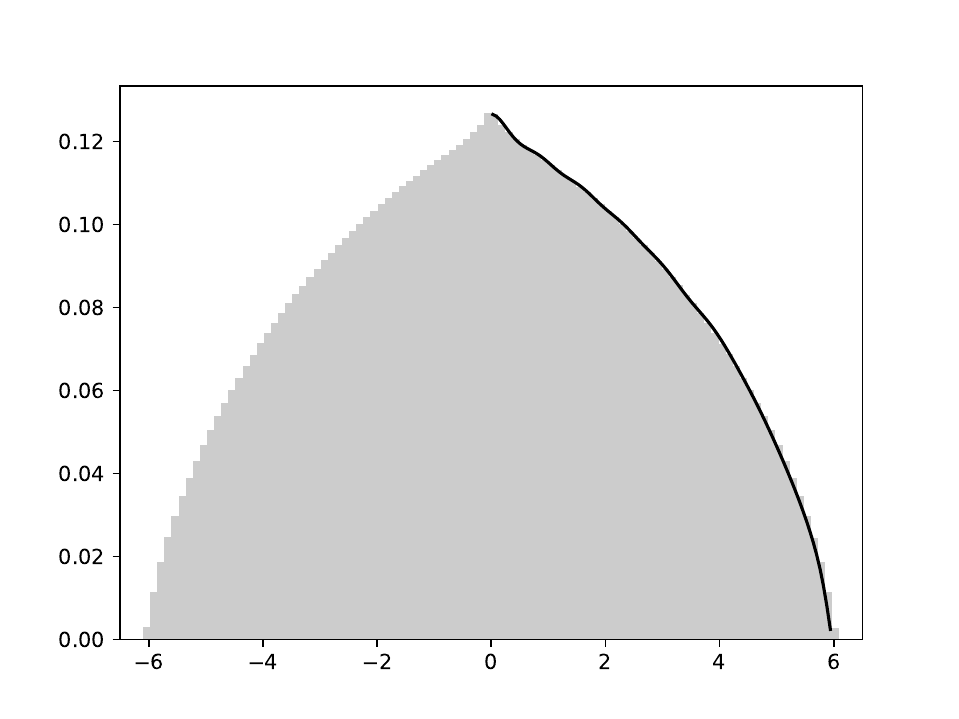}
\begin{picture}(0,0)
\put(-435,140){$p_{\mathrm{adj}}$}
\put(-345,0){$\la$}
\put(-300,140){$c=15$}
\put(-200,140){$p_{\mathrm{adj}}$}
\put(-107,0){$\la$}
\put(-65,140){$c=8$}
\end{picture}\vspace{2mm}
\caption{Solutions of the collocation problem (\ref{collocadj}) for \ER graph adjacency matrices at $J=10$, $\gamma=1$, converted to eigenvalue distribution estimates using (\ref{padjbeta}) and plotted as solid black lines for {\bf (left)} $c=15$ and {\bf (right)} $c=8$. As the distributions are reflection-symmetric, only the $\la>0$ part is plotted explicitly. The grey shaded areas represent empirical eigenvalue density histograms obtained from a sample of 1000 \ER graphs with 10000 vertices each at the corresponding values of $c$. Tiny undulations are seen near the center of the $c=8$ plot. These numerical artifacts become more problematic at smaller values of $c$.}
\label{figadj}
\end{figure}

\noindent To check the performance, we program this in Python (a sample script is provided in Appendix~\ref{appadj}) using the {\tt root} function of
the SciPy library \cite{SciPy} to solve (\ref{collocadj}) numerically with $\ga=1$ and the initial seed for the root search taken as $\be_0=1$, $\be_{j>0}=0$ at the first $z$-point, with the previous solutions reused as the seeds for each subsequent $z$-point. 

The results of our simulations are presented in Fig.~\ref{figadj}. It can be seen that the performance is perfect at $c=15$, which is shared by any higher values of $c$. At $c=8$ tiny undulations are seen near the center. These undulations become much more dramatic at lower $c$, disrupting the precision of the algorithm performance, though the overall shape of the distribution remains qualitatively correct. We shall comment in the conclusions on possible strategies for stabilizing the numerical performance at lower values of $c$.


\subsection{Ordinary graph Laplacians}

Graph Laplacians are of central importance for transport phenomena on networks. We start with the ordinary graph Laplacian
that controls, in particular, random walks on graphs where there is a fixed transition rate per edge per unit time \cite{laplace}. In terms of the adjacency matrix $\mathbf{A}$, drawn for \ER graphs from the distribution (\ref{adjdef}), this Laplacian is defined as follows: we first construct the diagonal degree matrix $\mathbf{D}$ whose diagonal entries are $D_{ii}\equiv \sum_j A_{ij}$, and then the ordinary Laplacian $\mathbf{L}$ is defined by
\beq\label{Ldef}
\mathbf{L}=\mathbf{D}-\mathbf{A}.
\eeq
Unlike the adjacency matrix, this matrix is positive semi-definite.

An analytic theory of the ordinary Laplacian along the lines of section~\ref{seqSFT} was developed in \cite{laplace}. One has to solve the Hammerstein equation
\beq\label{HammL}
g(\rho)=-ic\left\{e^{i\rho}-1-\rho\,e^{i\rho}\int_0^\infty\hspace{-3mm}d\rho'\,\frac{J_1(2\sqrt{\rho\rho'})}{\sqrt{\rho\rho'}}e^{ig(\rho')-i(z-1)\rho'}\right\}
\eeq
that replaces (\ref{sddlrho}) derived for adjacency matrices. The eigenvalue density of (\ref{Ldef}) is recovered from the solution as
\beq\label{extractpL}
p_L(\lambda)=\frac1\pi\,\Re\int_0^\infty \!\!\!d\rho\, e^{ig(\rho)-iz\rho}\Big|_{z=\la}=\frac1{\pi c}\Im[\del_\rho g(\rho,z)]\Big|_{\rho=0,\,\,z=\lambda+1},
\eeq
where we have restored, based on the principles
outlined in section~\ref{seqSFT}, the exact probability density normalization, ignored in the analytic formulas of \cite{laplace}. To get to the last representation, one uses the $\rho$-derivative of (\ref{HammL}) at $\rho=0$. Equations of the form (\ref{HammL}) date back to \cite{BR} where the derivations were based on replica methods.
They have also been derived, together with equations like (\ref{sddlrho}) for adjacency matrices, in \cite{Khetal} where the analysis based
on the methods of moments. The latter strategy has the advantage of being mathematically rigorous, but requires a genuine {\it tour de force}
of combinatorial accounting and resummations.

As in the case of adjacency matrices, it is convenient to implement a $c$-dependent rescaling to make the relevant ranges of variables depend only weakly on $c$. We choose this rescaling as
\beq\label{scaleL}
z=c+1+{\tilde z}{\sqrt{c}},\qquad \rho=\frac{\tilde \rho}{\sqrt{c}},\qquad g(\rho(\tilde\rho),z(\tilde z))=\tilde g(\tilde\rho,\tilde z)+\tilde\rho\sqrt{c}.
\eeq
The resulting equation, with tildes dropped, is
\beq\label{HammLscale}
g(\rho)=-ic\left(e^{i\rho/\sqrt{c}}-1-\frac{i\rho}{\sqrt{c}}\right)+i\rho\,e^{i\rho/\sqrt{c}}\int_0^\infty\hspace{-3mm}d\rho'\,\frac{J_1(2\sqrt{\rho\rho'/c})}{\sqrt{\rho\rho'/c\,}}e^{ig(\rho')-iz\rho'}.
\eeq
At large $c$, this equation reduces to $g(\rho)=i\rho^2/2+i\rho\int d\rho'\,e^{ig(\rho')-iz\rho'}$ \cite{laplace,Fyodorov,laplmoment}, which describes a `free convolution of the Wigner semicircle and a Gaussian' in the language of free probability theory \cite{freep}.
To represent (\ref{HammLscale}) in the form (\ref{Hammdef}), we choose
$$
s(\rho)=-ic\left(e^{i\rho/\sqrt{c}}-1-\frac{i\rho}{\sqrt{c}}\right), \quad K(\rho,\rho')=i\rho\,e^{i\rho/\sqrt{c}}\,\frac{J_1(2\sqrt{\rho\rho'/c})}{\sqrt{\rho\rho'/c\,}},
\quad F(\rho,g(\rho))=e^{ig(\rho)-iz\rho}. 
$$
We then follow the Kumar-Sloan strategy for implementing collocation described in section~\ref{seqHamm}, and specifically apply the expansion (\ref{expexp}).

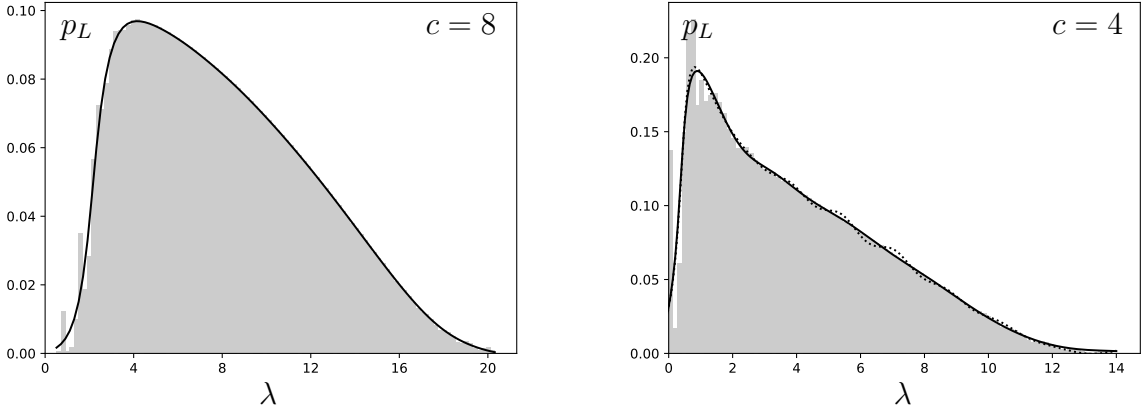
\begin{figure}[t]\vspace{-10mm}
\centering
\includegraphics[width = 0.49\linewidth]{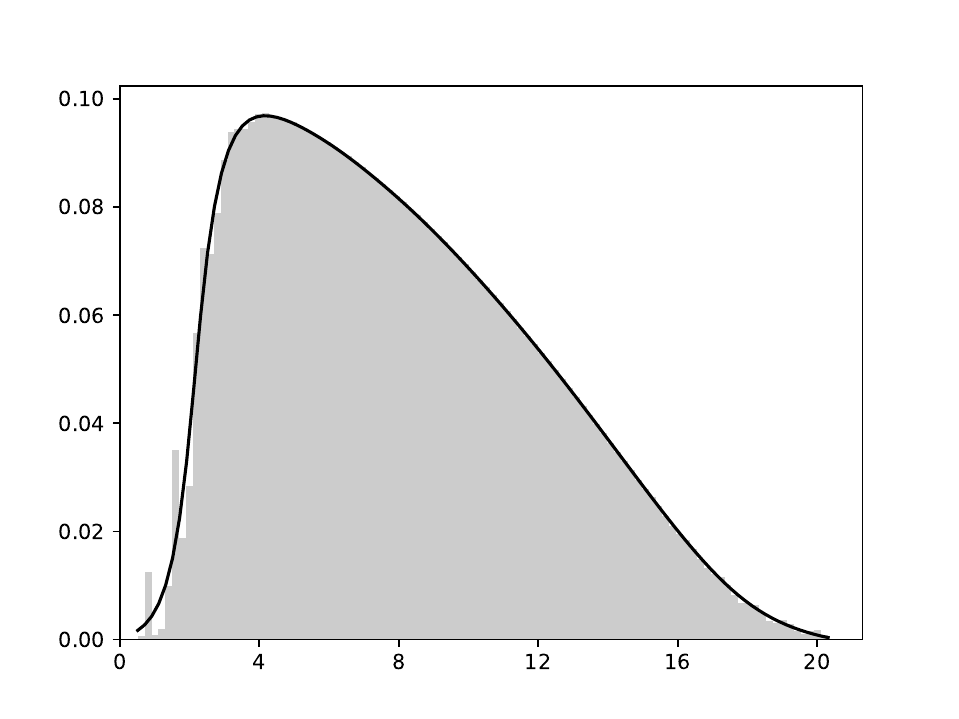}\hspace{2mm}\includegraphics[width = 0.49\linewidth]{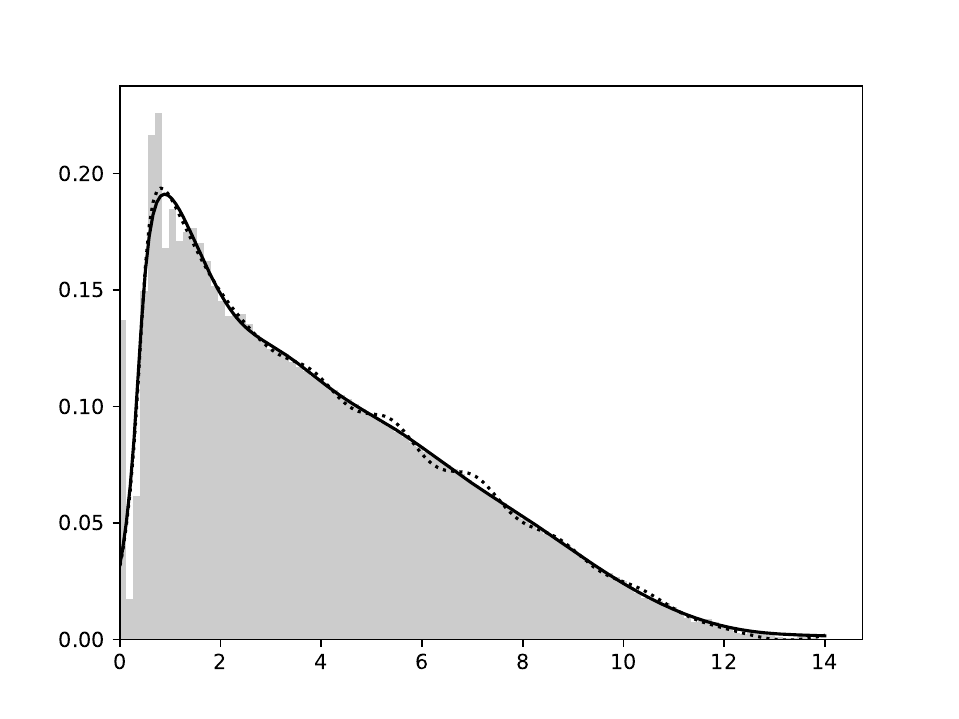}
\begin{picture}(0,0)
\put(-435,140){$p_L$}
\put(-360,0){$\la$}
\put(-297,140){$c=8$}
\put(-200,140){$p_L$}
\put(-120,0){$\la$}
\put(-60,140){$c=4$}
\end{picture}\vspace{2mm}
\caption{Results for the eigenvalue density of the ordinary graph Laplacian plotted as solid black lines: {\bf (left)} at $c=8$ using the Laguerre collocation problem (\ref{collocidL}) with $J=10$, $\gamma=5$, from which the eigenvalue density is extracted via (\ref{pL}), {\bf (right)} at $c=4$ using the general Kumar-Sloan collocation (\ref{colloc}) with the half-range Hermite expansion (\ref{hrHermite}) at $J=10$, followed by the integration in (\ref{extractpL}). 
Additionally, on the right panel, we add a dotted curve representing the result of solving (\ref{collocidL}) with $J=10$, $\gamma=5$ at $c=4$, showing more spurious undulations than the half-range Hermite method.
The grey shaded areas represent empirical eigenvalue density histograms obtained from a sample of 1000 \ER graphs with 10000 vertices each at the corresponding values of $c$.}
\label{figL}
\end{figure}

Since the integral over $\rho'$ in (\ref{HammLscale}) is exactly the same as in (\ref{adjrescaled}), we can rely on the same strategy as under (\ref{Besselid}) to explicitly evaluate all the integrals under the ansatz (\ref{expexp}). This yields, instead of (\ref{grec}), 
\beq\label{grecL}
g(\rho)=-ic\left(e^{i\rho/\sqrt{c}}-1-\frac{i\rho}{\sqrt{c}}\right)
+ic e^{i\rho/\sqrt{c}}\sum_{m=0}^{J}\be_m\left(1- e^{-\rho/\ga c}\sum_{n=0}^{m}\frac{(\rho/\ga c)^n}{ n!}\right),
\eeq
and leads to the following collocation problem:
\begin{align}\label{collocidL}
\sum_{j=0}^J \beta_j L_j(x_k)=\exp\Bigg[&x_k-\frac{izx_k}{\ga}+c\left(e^{ix_k/\ga\sqrt{c}}-1-\frac{ix_k}{\ga\sqrt{c}}\right)\\
&-ce^{ix_k/\ga\sqrt{c}}\sum_{m=0}^{J}\be_m\left(1- e^{-x_k/\ga^2 c}\sum_{n=0}^{m}\frac{(x_k/\ga^2 c)^n}{ n!}\right)\Bigg],\nn
\end{align}
where $x_k$ are the roots of $L_{J+1}$.
From the solution of this algebraic system, one recovers via (\ref{extractpL}), (\ref{scaleL}) and (\ref{grecL})
\begin{align}
&p_L(c+1+{\tilde z}{\sqrt{c}})=\frac{1}{\pi\sqrt{c}}\Re\int_0^\infty \!\!\!d\tilde\rho\, e^{i\tilde g(\tilde\rho)+i\tilde\rho\sqrt{c}-i(c+1+{\tilde z}{\sqrt{c}})\tilde\rho/\sqrt{c}}=\frac{1}{\gamma\pi\sqrt{c}}\Re\int_0^\infty \!\!\!dx\nn\\
&\hspace{1.5cm}\times \sum_{j=0}^J \beta_j L_j(x) \,e^{-(1+i/\ga\sqrt{c})x}=\frac{1}{\gamma\pi\sqrt{c}}\Re\int_0^\infty \!\!\!dx\, \sum_{j=0}^J \beta_j \sum_{n=0}^j {j\choose n}\frac{(-x)^n}{n!} \,e^{-(1+i/\ga\sqrt{c})x}\nn\\
&\hspace{1.5cm}=\frac{1}{\gamma\pi\sqrt{c}}\Re \sum_{j=0}^J \beta_j \sum_{n=0}^j {j\choose n}\frac{(-1)^n}{(1+i/\ga\sqrt{c})^{n+1}}=
\frac{1}{\pi}\Im \sum_{j=0}^J \frac{\hspace{-5mm}\beta_j(\tilde z)}{\left(1-i\ga\sqrt{c}\right)^{j+1}}.\label{pL}
\end{align}
We have performed numerical simulations based on (\ref{collocidL}) and (\ref{pL}) and observed that they reproduce very well the empirical distribution at $c\gtrsim 8$, though increasing the expansion parameter $\gamma$ becomes necessary towards the lower end of this range to ensure numerical convergence. At smaller values of $c$, the performance is less stable, especially in regions away from the main peak, where spurious oscillations develop. The performance at $c=8$ is seen on the left panel of Fig.~\ref{figL}. It captures the smooth descending part of the distribution perfectly, but is less accurate to the left of the main peak, where rough textures develop as $c$ tends to the `percolation threshold' at $c=1$, even though the overall shape is reproduced correctly.

One can understand intuitively the deteriorating performance of the expansion scheme (\ref{expexp}) away from the main peak. Indeed, there, the first derivative of $g(\rho)$ at $\rho=0$ becomes small and $e^{ig-iz\rho}$ is dominated by the Gaussian envelope $e^{-\rho^2/2}$, clearly visible \cite{laplace} in the large $c$ limit of ($\ref{HammLscale})$. If we expand as in (\ref{expexp}), this Gaussian envelope is approximated by a polynomial, which is clearly prone to instabilities due to the large $\rho$ growth of polynomials and large $\rho$ decay of $e^{-\rho^2/2}$. To remedy for this problem, we can replace (\ref{expexp}) by an alternative expansion
\beq\label{hrHermite}
e^{ig(\rho)-iz\rho}=\sum_{j=0}^J \beta_j H_j(\rho) \,e^{-\rho^2/2},
\eeq
where $H_j$ are the half-range Hermite polynomials \cite{hr} orthogonal with respect to the inner product $\int_0^\infty \hspace{-1mm} d\rho\, e^{-\rho^2}H_i(\rho) H_j(\rho) $. The price to pay is that there are no longer simple and nice analytic formulas for $H_j$, but we can still process all the integrals numerically and run the generic formulation of Kumar-Sloan collocation given by (\ref{colloc}). We have observed that this approach performs better at smaller $c$. (A sample script is given in Appendix~\ref{appherm}.) In the right panel of Fig.~\ref{figL}, the performance of this scheme at the low value $c=4$ is shown. The smooth heavy tail of the distribution is captured very accurately, while the rough part around the ascent and the peak are reproduced less well, though the overall shape is correct.

At larger values of $c$, both approaches quickly become very accurate, and the empirical distribution shape is effectively indistinguishable from our numerical solutions of the Hammerstein equation (\ref{HammL}) already at c=15.


\subsection{Normalized graph Laplacians}

The normalized Laplacian is defined, in the notation of (\ref{Ldef}), by
\beq\label{normLa}
 \pmb{\mathcal{L}}\equiv\mathbf{D}^{-1/2}(\mathbf{D}-\mathbf{A})\mathbf{D}^{-1/2},
\eeq
This Laplacian controls, in particular, random walks where there is a fixed rate per unit time to leave the present vertex and jump with equal probability to one of its neighbors \cite{laplace}. (In the above defintion, $\mathbf{D}^{-1/2}$ is understood as the square root of the pseudoinverse of $\mathbf{D}$.)

The spectral theory of normalized Laplacians of \ER graphs was developed in \cite{laplace}. It results in the following Hammerstein equation:
\beq\label{sddlLn}
g(\rho)=-ic\left(e^{i(1-z)\rho}-1-\rho\,e^{i(1-z)\rho}\int_0^\infty\hspace{-3mm}d\rho'\,\frac{J_1(2\sqrt{\rho\rho'})}{\sqrt{\rho\rho'}}\,e^{ig(\rho')+i(1-z)\rho'}\right).
\eeq
(Curiously, the same equation was recently found to control the probabilities of first return times of random walks
on \ER graphs \cite{walkongraph}.)
From the solution of this equation, the eigenvalue density can be recovered in the following form, with the normalization given explicitly:
\beq\label{pLint}
p_{\mathcal{L}}(\lambda)=\frac1\pi \Re\int_0^\infty\hspace{-3mm} d\rho\,[c+i g(\rho)]\,e^{ig(\rho)}\Big|_{z=\lambda}.
\eeq
As in our previous treatments, it is convenient to introduce rescalings that minimize the dependence of the contributing ranges of variables on $c$. This is accomplished with
\beq\label{scalingLn}
z=1+\frac{\tilde z\sqrt{c}}{c+1},\qquad \rho=\frac{\tilde{\rho}}{\sqrt{c}},\qquad g(\rho(\tilde\rho),z(\tilde z))=\tilde g(\tilde\rho,\tilde z)-\tilde z\tilde\rho\frac{c}{c+1}.
\eeq
Dropping all tildes, we then get
\beq\label{sddlLnscaled}
g(\rho)=-ic\left(e^{\frac{-iz\rho}{c+1}}-1+\frac{iz\rho}{c+1}\right)+i\rho\,e^{\frac{-iz\rho}{c+1}}\int_0^\infty\hspace{-3mm}d\rho'\,\frac{J_1(2\sqrt{\rho\rho'/c})}{\sqrt{\rho\rho'/c}}\,e^{ig(\rho')-iz\rho'}.
\eeq
The $c\to\infty$ limit $g(\rho) = i\rho\int d\rho' \,e^{ig(\rho')-iz\rho'}$ is the same as for adjacency matrices, defining a Wigner semicircle \cite{laplace}.
For the Kumar-Sloan collocation, we identify $F(\rho,g(\rho))=e^{ig(\rho)-iz\rho}$ and
\beq
s(\rho)=-ic\left(e^{\frac{-iz\rho}{c+1}}-1+\frac{iz\rho}{c+1}\right), \quad K(\rho,\rho')=i\rho\,e^{\frac{-iz\rho}{c+1}}\,\frac{J_1(2\sqrt{\rho\rho'/c})}{\sqrt{\rho\rho'/c\,}}. 
\eeq
The integral over $\rho'$ in (\ref{sddlLnscaled}) is once again completely identical to (\ref{adjrescaled}), so we adopt the expansion (\ref{expexp}) and apply the same processing as in section~\ref{subseqadj} to obtain
\begin{align}\label{gLn}
g(\rho)=&-ic\left(e^{\frac{-iz\rho}{c+1}}-1+\frac{iz\rho}{c+1}\right)+ice^{\frac{-iz\rho}{c+1}}\sum_{m=0}^{J}\be_m\left(1- e^{-\rho/\ga c}\sum_{n=0}^{m}\frac{(\rho/\ga c)^n}{ n!}\right).
\end{align}
The collocation problem, analogous to (\ref{collocadj}) for adjacency matrices,
then becomes
\begin{align}\label{collocidLn}
\sum_{j=0}^J \beta_j L_j(x_k)& =\exp\Bigg[x_k-\frac{izx_k}{\ga}+c\Bigg(e^{\frac{-izx_k}{\ga(c+1)}}-1+\frac{izx_k}{\ga(c+1)}\Bigg)\\
&\hspace{3cm}-ce^{\frac{-izx_k}{\ga(c+1)}}\sum_{m=0}^{J}\be_m\Bigg(1- e^{-x_k/\ga^2 c}\sum_{n=0}^{m}\frac{(x_k/\ga^2 c)^n}{ n!}\Bigg)\Bigg],\nn
\end{align}
where the collocation points $x_k$ are the roots of $L_{J+1}$.
After these equations have been solved numerically,
one must reconstruct $e^{ig}$ from (\ref{expexp}), $g$ from (\ref{gLn}), undo the rescaling (\ref{scalingLn}) and evaluate (\ref{pLint}). In this way, we obtain:
\begin{align}\label{pLn}
&p_{\mathcal{L}}(\lambda)=\frac{\sqrt{c}}{\pi\gamma} \Re\int_0^\infty\hspace{-3mm} dx\,\Bigg[1-\sum_{m=0}^{J}\be_m\Big(1- e^{-x/\ga^2 c}\sum_{n=0}^{m}\frac{(x/\ga^2 c)^n}{ n!}\Big)\Bigg]\sum_{j=0}^J \beta_j L_j(x) \,e^{-x}\\
&=\frac{\sqrt{c}}{\pi\gamma} \Re\Bigg[\beta_0\Big(1-\sum_{m=0}^J\beta_m\Big)+\sum_{j,m=0}^J \beta_m\beta_j\sum_{n=0}^{m}\sum_{k=0}^j{j\choose k}\frac{(-1)^k}{n!k!(\ga^2 c)^n}\int_0^\infty\hspace{-3mm} dx\, x^{n+k}e^{-x(1+1/\ga^2c)}\Bigg]\nn\\
&=\frac{\sqrt{c}}{\pi\gamma} \Re\Bigg[\beta_0\Big(1-\sum_{m=0}^J\beta_m\Big)-\sum_{j,m=0}^J \beta_m\beta_j\sum_{n=0}^{m}\sum_{k=0}^j{j\choose k}{n+k\choose k}\frac{(-\gamma^2c)^{k+1}}{(\ga^2c+1)^{n+k+1}}\Bigg]\Bigg|_{z=\frac{(\lambda-1)(c+1)}{\sqrt{c}}}.\nn
\end{align}
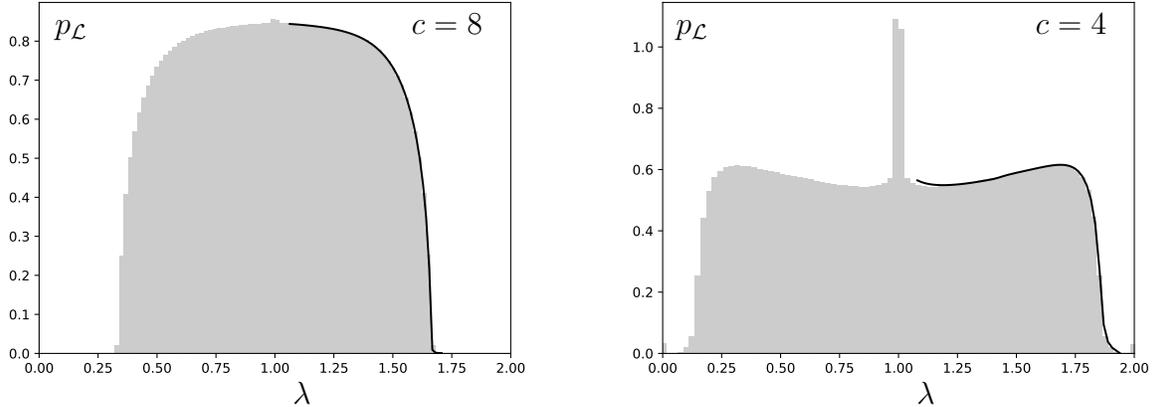
\begin{figure}[t]\vspace{-10mm}
\centering
\includegraphics[width = 0.49\linewidth]{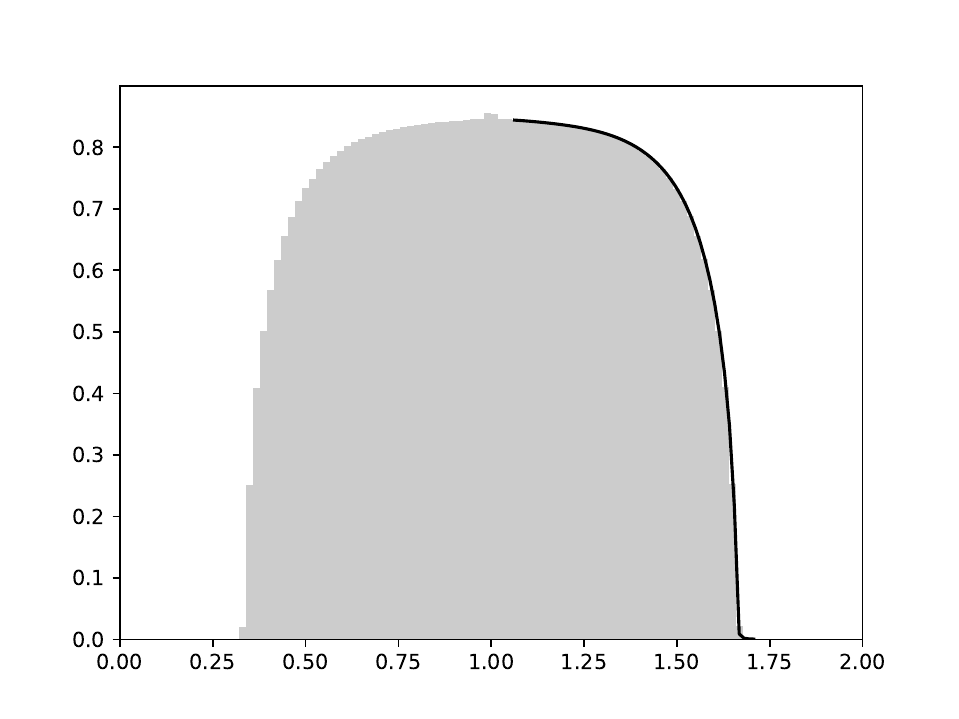}\hspace{2mm}\includegraphics[width = 0.49\linewidth]{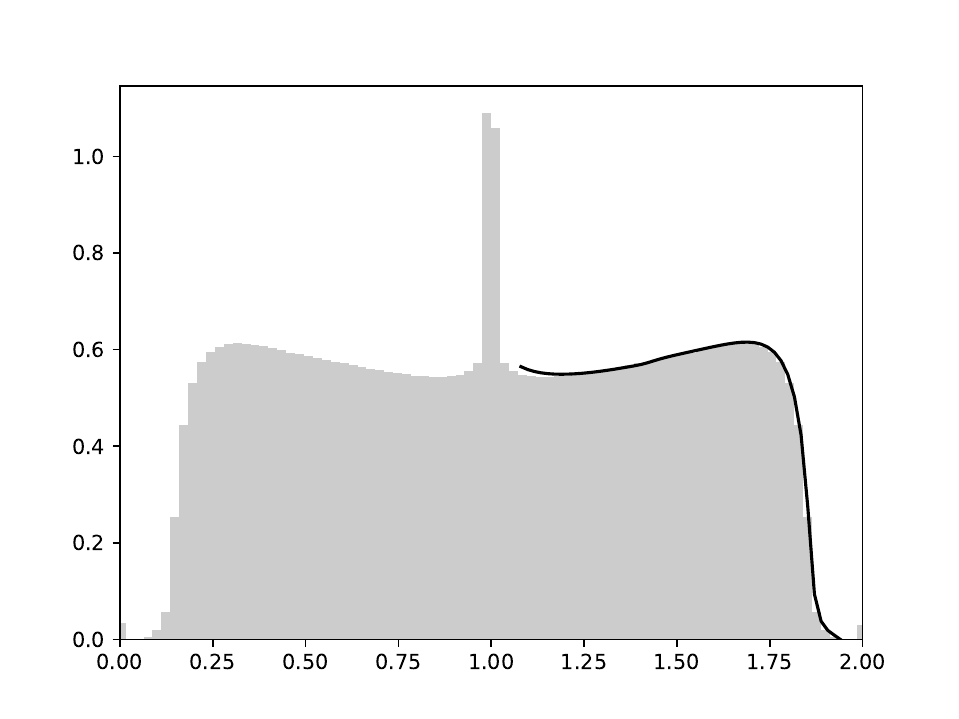}
\begin{picture}(0,0)
\put(-435,140){$p_{\mathcal{L}}$}
\put(-345,0){$\la$}
\put(-300,140){$c=8$}
\put(-200,140){$p_{\mathcal{L}}$}
\put(-109,0){$\la$}
\put(-64,140){$c=4$}
\end{picture}\vspace{2mm}
\caption{Solutions of the collocation problem (\ref{collocidLn}) for the normalized Laplacians of \ER graphs at $J=9$, $\gamma=2$, converted to eigenvalue distribution estimates using (\ref{pLn}) and plotted as solid black lines for {\bf (left)} $c=8$ and {\bf (right)} $c=4$. As the distributions are reflection-symmetric, only the $\la>1$ part is plotted explicitly. We excise small regions near $\la=1$ where the convergence of our collocation scheme is compromised by the presence of sharp spikes in the actual eigenvalue distribution. (There are generically $\de$-function spikes at $\la=1$ that become more and more prominent at small $c$. We comment on them further in the conclusions.) The grey shaded areas represent empirical eigenvalue density histograms obtained from a sample of 1000 \ER graphs with 10000 vertices each at the corresponding values of $c$.}
\label{figLn}
\end{figure}

\noindent We display the output of solving (\ref{collocidLn}), followed by an application of (\ref{pLn}) in Fig.~\ref{figLn}. The performance of this algorithm remains stable and successful down to rather low values $c=8$ and 4. Convergence is more problematic near the origin at low $c$, where a large abrupt spike develops, hence we excise these regions. In the bulk of the distribution, the numerics based on (\ref{collocidLn}) and (\ref{pLn}) gives very convincing results.


\section{Outlook}\label{seqfin}

We have applied a systematic treatment of nonlinear integral equations of the Hammerstein type to recover the spectra of sparse random matrices
associated to \ER graphs with an asymptotically large number of vertices of average degree $c$. Our treatment is practically exact for $c$ larger than 15 or so, and remains rather accurate down to $c$ around 8. As $c$ is lowered further, approaching the percolation threshold at $c=1$, challenges in numerical performance are met, though for graph Laplacians, we still obtain very reasonable output for $c$ as low as 4. We hope that further improvements in numerical methods for solving Hammerstein equations will lead to better performance in the low-$c$ region.

The approach we have taken here is very complementary to the widely explored `cavity' and `population dynamics' methods for reconstructing sparse graph spectra \cite{kuehn, cavity,suscaetal}. In those approaches, stochastic simulations are designed whose stationary distributions contain information
on eigenvalue densities. By contrast, our numerics is fully deterministic and consists in solving the integral equations (\ref{sddlrho}), (\ref{HammL}) and (\ref{sddlLn}) using collocation methods. The implementation is very economical in its computational costs: numerical solutions of the Hammerstein equations in all of our examples (the solid black lines in the plots) take seconds to obtain on an ordinary personal computer. We feel that it would be very interesting
to apply this technology to more sophisticated sparse matrix problems, for example, those arising in the theory of amorphous solids \cite{amorph} or for non-Hermitian matrices \cite{Baron,nherm,randgen}.

The main challenges met by our numerical algorithms are either in the regions where the eigenvalue density becomes very small, 
or at low values of $c$, where it is known from empirical diagonalization of large matrices that the regularity of the eigenvalue density deteriorates dramatically.
We provide below preliminary comments on the origin of these issues and potential pathways to their solution.

The poor performance outside the main support of the eigenvalue density can be easily understood qualitatively. The eigenvalue density is related to the  imaginary part of the $\rho$-derivative of $g$ at the origin, as in (\ref{extractpL}). When the eigenvalue density is small, the imaginary part of $g$ grows slowly with $\rho$, and hence $e ^{ig}$ decreases slowly with $\rho$. As a result, a large range of $\rho'$ becomes relevant in the integrals on the right-hand side of the Hammerstein equations, with complicated oscillatory behaviors, and one should not expect that such integrals will be easily approximated using interpolating polynomials. The situation is significantly better for ordinary Laplacians, where $e^{ig}$ includes a Gaussian envelope $e^{-\rho^2/2}$, independent of the eigenvalue density, and this tames the integrals over $\rho'$ when we use the half-range Hermite polynomial expansion adapted to this Gaussian envelope. (We note that the other cases also have similar Gaussian envelopes, but the width is of order $\sqrt{c}$ rather than of order 1. This may provide a viable alternative strategy at small $c$, again, moving away from simple Laguerre expansions.) We feel that a general scheme to handle the equations outside the main support of the eigenvalue density should exist, possibly by rearranging the integrals in the complex $\rho$-plane, but we have not been able to formulate a working recipe. This problem is of somewhat marginal importance for the eigenvalue distribution topic, since one is mostly interested in the shape of the eigenvalue distributions where they are nonvanishing, and not in the regions where they (almost) vanish. By contrast, for applications of Hammerstein equations to random walks on random graphs, as in \cite{walkongraph}, the behavior of $g$ near $z=\infty$, very far away from the main support of the eigenvalue density, is of primary importance.

Using Laguerre polynomial expansions, we have been able to evaluate all the integrals in the Hammerstein equations with Bessel kernels exactly,
obtaining explicit systems of algebraic equations (\ref{collocadj}), (\ref{collocidL}) and (\ref{collocidLn}). This both provides for very fast numerical performance and creates a comfortable environment for more detailed mathematical analysis of convergence, approximation accuracy, etc.
Where it is advantageous to use a different kind of expansion, as in our half-range Hermite treatment of the ordinary graph Laplacians,
the integrals can still be effectively handled numerically following the general Kumar-Sloan scheme (\ref{colloc}).

As to improving the performance of our numerics at lower $c$, the issue seems to be that the shape of various functional dependences becomes more complicated at lower $c$, and our simple polynomial expansions fail with capturing it. One could hope that increasing the polynomial order $J$ in (\ref{expexp}) would help, but it is not straightforward for two reasons. First, as we go to higher $J$, we would need to solve nonlinear systems of equations similar to (\ref{collocadj}) in higher and higher numbers of dimension, where naive root searches become less and less reliable. Perhaps this could be tackled by first running the algorithm at a lower value of $J$ and then reusing the output as seeds for the root search at higher values of $J$. More dramatically, however,
the interpolating projector we have used with great efficiency at low $J$ is likely to be flawed at higher $J$. It is well-known that naively increasing the order of interpolation does not necessarily result in better approximation \cite{runge}. There could certainly be more refined choices for the projector $\Prj $ in (\ref{projexp}) and other similar equations. A naive option is to rely on the infinite radius of convergence of the Taylor series for exponentials and employ a Taylor expansion on the right-hand side of (\ref{projexp}). This, however, would require working at a very high arithmetic precision, since the convergence of Taylor expansions for exponentials requires massive numerical cancellations. There likely exist better choices for $\Prj $ that we leave for future investigations, also hoping for practical input from mathematicians specializing in numerical methods.
Uniform approximations of functions by polynomials on infinite intervals in the presence of exponential suppression have been discussed systematically
in the literature \cite{weighted}.

In our numerical work, we have treated the expansion parameters such as $\ga$ in (\ref{expexp}) and the initial seed for multidimensional root searches as free parameters chosen empirically to ensure satisfactory performance. In the cases we considered, simple straightforward choices sufficed to obtain the plots given in the paper. It could be good to develop a better approach where these parameters are bootstrapped toward good values, either by analyzing the integral equations at a given value of $z$, or by reusing the previously found solutions at other values of $z$ (the latter is actually the approach we took for the initial seed).

Nonlinear algebraic equations like (\ref{collocadj}), (\ref{collocidL}) and (\ref{collocidLn}) are generally expected to have multiple solutions.
Only one of these solutions corresponds to the actual eigenvalue density curve, while the others are spurious. We could see that deforming the parameters of
our expansion schemes (for example, making $\ga$ small) or the initial seed far from the optimal values we found leads to such spurious convergence,
predicting a completely wrong eigenvalue density curve. For the concrete cases we considered, we indicated simple natural choices of the numerical
implementation parameters that ensure convergence to the correct solution. It would be nice to have more of a theory of the solutions
of (\ref{collocadj}), (\ref{collocidL}) and (\ref{collocidLn}) and their basins of attraction relative to root search algorithm, especially given that the
equations are provided explicitly, with all the integrals evaluated in a closed form. Uniqueness of solutions of Hammerstein equations
(before converting them to approximate collocation schemes) has been discussed extensively in the mathematical literature. 

One striking feature of the eigenvalue distributions of sparse matrices at lower values of $c$ is the emergence of sharp peaks \cite{kuehn,BG,delta}.
These peaks are understood qualitatively as coming from small disconnected components. For example, isolated vertices of degree 0 contribute zero eigenvalues to the adjacency matrix. Since there are on average $e^{-c}N$ vertices of degree 0, one expects a $e^{-c}\de(\lambda)$ peak in the eigenvalue distribution. Similarly, graph Laplacians develop zero eigenvalues for each disconnected component of the graph. There are further peaks at nonzero values of $\lambda$. We suspect that they come from graph components that are only connected to the rest of the graph by one edge, and other peculiar arrangements (similar structures play a significant role in the distribution of resistance distances \cite{resdist}). These sharp peaks are, incidentally, reflected differently in the empirical histograms and in the numerical solutions of the integral equations appearing in our plots. Numerical histograms, in the limit of large samples, always show the integrals of the eigenvalue density over single-bin intervals. The numerical solutions are evaluated at a single $\la$-point, except that the truncated approximations of the original integral equation, as in (\ref{expexp}), are likely to resolve the infinitely thin $\de$-function peaks somewhat.

The sharp peaks could be of significant interest physically, depending on the setting. For example, where random matrix spectra represent vibrational modes, the peaks in the eigenvalue density will signify strongly resonant points in the spectrum. The $\de$-function singularities in the eigenvalue distribution
should translate to singular behaviors of $g(z;\rho=0)$ at some special values of $z$. It would be very nice to develop a systematic theory of these sharp peaks, and use this analytic understanding to subtract that part within the equations of $g$, so that the target functions for numerical work become more smooth and the numerical methods become more stable.

\section*{Acknowledgments}

OE is supported by Thailand NSRF via PMU-B, grant number B13F670063.  
We acknowledge the use of computational resources of LPTMS at Universit\'e Paris-Saclay for generating and diagonalizing large samples of random matrices.

\appendix

\section{A more general expansion in terms of Laguerre polynomials}\label{appexp}

Instead of (\ref{expexp}), we could have employed a more general expansion
\beq\label{expexpab}
e^{ig(\rho)-iz\rho}=\sum_{j=0}^J \beta_j L_j(a\rho) \,e^{-b\rho},
\eeq
where the scalings of the argument of the Laguerre polynomial and the exponential are different. This does not immediately yield significant improvements over the case $a=b=\ga$ treated in the main text, and the formulas become more bulky. These formulas, however, may become useful for other future applications, so we give a summary here on how to treat the adjacency matrix equation (\ref{adjrescaled}) with the expansion (\ref{expexpab}).

The integrals can in general be evaluated using the relation of the Bessel convolutions -- that is, Hankel transforms -- to 2d Fourier transforms. Specifically, we write $J_1(\rho)=-\del_\rho J_0(\rho) =-\pi^{-1}\del_\rho \int_0^\pi d\ph \,e^{-i\rho\cos\ph}$. Then, for any $f(\rho)$ that decays at infinity,
\begin{align}
&\rho\int_0^\infty \hspace{-2mm} d\rho'\, \frac{J_1(2\sqrt{\rho\rho'/c})}{\sqrt{\rho\rho'/c\,}}f(\rho') =-\frac{c}\pi\int_0^\infty \hspace{-2mm}d\rho' \int_0^\pi\hspace{-2mm} d\ph\,
\del_{\rho'}e^{-2i\sqrt{\rho\rho'/c}\cos\ph}f(\rho')=cf(0)\label{xyint}\\
&+\frac{c}\pi\int_0^\infty \hspace{-2mm}d\rho' \int_0^\pi\hspace{-2mm} d\ph\,
e^{-2i\sqrt{\rho\rho'/c}\cos\ph}\del_{\rho'}f(\rho')=\frac{2c}\pi\int_{-\infty}^\infty \hspace{-2mm}dx \int_0^\infty\hspace{-2mm} dy\,
e^{-2ix\sqrt{\rho/c}}\del_{\rho'}f(\rho')\Big|_{\rho'=x^2+y^2}+cf(0),\nn
\end{align}
where we have introduced the Cartesian coordinates $x\equiv \sqrt{\rho'}\cos\ph$, $y\equiv \sqrt{\rho'}\sin\ph$.

If we start with the definition of Laguerre polynomials
\beq\label{Lagdef}
L_j(a\rho)e^{-b\rho}=e^{-b\rho} \sum_{n=0}^j {j\choose n}\frac{(-a\rho)^n}{n!},
\eeq
substitute $\rho=x^2+y^2$
and expand everything in terms of monomials of the form $x^{2k}y^{2l}$, the integrals are evaluated term-by-term using the two elementary Gaussian quadratures:
\begin{align}
&\int_{-\infty}^\infty \hspace{-2mm}x^{2n}e^{2ix\sqrt{\rho/c}} e^{-bx^2}dx=e^{-\rho/bc}\int_{-\infty}^\infty \hspace{-2mm}(\tilde x+i\sqrt{\rho/c})^{2n} e^{-b\tilde x^2}d\tilde x\\
&=e^{-\rho/bc}\sum_{k=0}^n{2n\choose 2k}\left(\frac{-\rho}c\right)^{\!\!n-k}\int_{-\infty}^\infty \hspace{-2mm} \tilde x^{2k} e^{-b\tilde x^2}d\tilde x
=e^{-\rho/bc}\sqrt{\frac\pi{b}}\sum_{k=0}^n{2n\choose 2k}\left(\frac{-\rho}c\right)^{\!\!n-k}\frac{(2k-1)!!}{(2b)^k},\nn
\end{align}
and
\beq
\int_{0}^\infty \hspace{-2mm} y^{2m} e^{-by^2}dy=\sqrt{\frac\pi{b}}\,\frac{(2m-1)!!}{2(2b)^m},
\eeq
where one must define $(-1)!!=1$ to make the formulas work correctly.
Then,
\begin{align}
&\int_{-\infty}^\infty \hspace{-2mm}dx\int_0^\infty \hspace{-2mm}dy \,(x^2+y^2)^n e^{2ix\sqrt{\rho/c}} e^{-b(x^2+y^2)}
= \sum_{m=0}^n{n \choose m} \!\!\int_{-\infty}^\infty \hspace{-3mm}x^{2m}e^{2ix\sqrt{\rho/c}} e^{-bx^2}dx\!\int_{0}^\infty \hspace{-3mm} y^{2(n-m)} e^{-by^2}dy\nn\\
&\hspace{2cm}=\frac\pi{2b}e^{-\rho/bc} C^{\mathrm{aux}}_n(\rho),
\end{align}
where we have introduced the following auxiliary polynomials
\beq\label{Cauxdef}
C^{\mathrm{aux}}_n(\rho)\equiv\frac{1}{(2b)^{n}}\sum_{m=0}^n{n \choose m}(2(n-m)-1)!!\sum_{k=0}^m{2m\choose 2k}(2k-1)!!\left(\frac{-2b\rho}{c}\right)^{\!\!m-k}.
\eeq
With these ingredients, we can write for the integrals of the individual terms in (\ref{Lagdef}):
\begin{align}
&\rho\int_0^\infty \hspace{-2mm}d\rho'\, \frac{J_1(2\sqrt{\rho\rho'/c})}{\sqrt{\rho\rho'/c\,}} \,\rho'^n e^{-b\rho'}=
\frac{2c}{\pi}\!\int_{-\infty}^\infty \hspace{-3mm}dx\!\int_0^\infty \hspace{-3mm}dy \,e^{2ix\sqrt{\rho/c}} \, [n(x^2+y^2)^{n-1}\!-\!b(x^2+y^2)^n]e^{-b(x^2+y^2)}\nn\\
&\hspace{1cm}=ce^{-\rho/bc}\Big[\frac{n}bC^{\mathrm{aux}}_{n-1}(\rho)-C^{\mathrm{aux}}_n(\rho)\Big].
\end{align}
To make this formula work correctly at $n=0$, we must add $c$ due to the last term in (\ref{xyint}).
If we now impose (\ref{adjrescaled}) at collocation points $\rho=\rho_l$ with $l=0..J$ under the ansatz (\ref{expexpab}) and use the above formulas to evaluate the Bessel integral, keeping in mind that $C^{\mathrm{aux}}_0(\rho)=1$, the output is the following collocation problem
\beq\label{collocadjab}
\sum_{j=0}^J L_{jl}\beta_j =\exp\Bigg[(b-iz)\rho_l+c(e^{-\rho_l/bc}-1)\sum_{j=0}^J \beta_j+\sum_{j=1}^J C_{jl}\beta_j\Bigg],
\eeq 
with $\displaystyle L_{jl}\equiv \sum_{n=0}^j {j\choose n}\frac{(-a\rho_l)^n}{n!}$ and
\begin{align}
& C_{jl}\equiv ce^{-\rho_l/bc} \sum_{n=1}^j {j\choose n}\frac{(-a)^n}{n!}\Big[C^\mathrm{aux}_{n}(\rho_l)
-\frac{n}{b}C^\mathrm{aux}_{n-1}(\rho_l)\Big],\label{Ccollocadj}
\end{align}
with $C^\mathrm{aux}$ given by (\ref{Cauxdef}).
While the multiple sums may seem awkward, they merely encode the coefficients of explicit polynomials in $\rho$ and can be treated very effectively in numerical implementations. They also need to be evaluated only once.
Note that no approximations have been made in (\ref{collocadjab}-\ref{Ccollocadj}) beyond truncating the expansion (\ref{expexpab}) at $j=J$.

For the collocation points we choose  $\rho_l=x_l/a$. where $x_l$ are the roots $L_{J+1}$, so that $L_{J+1}(x_l)=0$. We then use the above collocation problem to find $\beta_j$ and hence $e^{ig(\rho)-iz\rho}$, and then the eigenvalue density can be recovered from (\ref{extractp}),
which gives
\beq\label{extractpab}
p_\mathrm{adj}(\lambda)=\frac1{\pi b}\,\Re\sum_{j=0}^J \beta_j \sum_{n=0}^j {j\choose n}(-{a}/{b})^n=\frac1{\pi b}\sum_{j=0}^J (1-{a}/{b})^j \,\Re\beta_j\Big|_{z=\la/\sqrt{c}}.
\eeq

\section{Python code for the adjacency spectrum}\label{appadj}

We provide a basic Python script that recovers the solution of the collocation problem (\ref{collocadj}) for adjacency matrices.
Related collocation problems (\ref{collocidL}) and (\ref{collocidLn}) for graph Laplacians can be treated by minimal modifications of this script.

\begin{verbatim}
import numpy as np
import scipy.optimize as opt
import matplotlib.pyplot as plt

def eqs(beta,z):
 betac=beta[:J]+1j*beta[J:]
 eqs=[sum([L[k,j]*betac[j] for j in range(J)])-np.exp(x[k]*(1-1j*z/gamma)
      -sum([C[k,j]*betac[j] for j in range(J)])) for k in range(J)]   
 return np.append(np.real(eqs),np.imag(eqs))
 
c=15
sqrtc=np.sqrt(c)
J=11  #corresponds to J+1 in the notation of the paper
gamma=1
zs=np.linspace(0.01,2.1,100)
x=np.polynomial.laguerre.lagroots([0]*J+[1])

fac=np.ones((J))
for i in range(1,J-1):
 fac[i+1]=(i+1)*fac[i]

C=np.zeros((J,J))
L=np.zeros((J,J))
for k in range(J):
 xgc=x[k]/(gamma**2*c)
 expxgc=np.exp(-xgc)
 for j in range(J):
  L[k,j]=np.polynomial.laguerre.Laguerre([0]*j+[1])(x[k])
  C[k,j]=c*(1-expxgc*sum([xgc**n/fac[n] for n in range(j+1)]))

p=[]
beta=np.array([1]+[0]*(2*J-1))   
for z in zs:
 sol=opt.root(eqs,beta,args=(z),tol=1e-12)
 beta=sol.x
 p.append(beta[0]/(gamma*np.pi*sqrtc))

plt.plot(zs*sqrtc,p)
plt.ylim(0)
plt.show()
\end{verbatim}

\section{Python code for the ordinary Laplacian via half-range Hermite polynomials}\label{appherm}

We provide a simple Python script that implements the half-range Hermite collocation for ordinary graph Laplacians. While we could evaluate some of the integrals here analytically in principle, the numerical evaluation works well for practical purposes.

\begin{verbatim}
import numpy as np
from scipy.optimize import root
from scipy.integrate import quad
from scipy.special import jv
import matplotlib.pyplot as plt

def hrHermites(n):
 polys=[np.poly1d([(np.sqrt(np.pi)/2)**(-0.5)])]
 for i in range(1,n+1):
  this_poly=np.poly1d([1]+[0]*i)
  for last_poly in polys:
   this_poly-=last_poly*quad(lambda y: this_poly(y)*last_poly(y)*np.exp(-y**2),
                                             0,np.inf)[0]
  this_poly=this_poly/np.sqrt(quad(lambda y: this_poly(y)**2*np.exp(-y**2),
                                             0,np.inf)[0])
  polys.append(this_poly)
 return polys
 
def eqs(beta,z):
 betac=beta[:J]+1j*beta[J:]
 eqs=[]
 for k in range(J):
  xg=1j*x[k]/sqrtc
  expxg=np.exp(xg)
  eqs.append(sum([H[k,j]*betac[j] for j in range(J)])-np.exp(x[k]**2/2-1j*x[k]*z
      +c*(expxg-1-xg)-expxg*sum([C[k,j]*betac[j] for j in range(J)])))
 return np.append(np.real(eqs),np.imag(eqs))

c=5
sqrtc=np.sqrt(c)
J=12
zs=np.linspace(-3,4,100)
hrH=hrHermites(J)
x=hrH[-1].r

C=np.zeros((J,J))
H=np.zeros((J,J))
for k in range(J):
 for j in range(J):
  H[k,j]=hrH[j](x[k])
  C[k,j]=sqrtc*quad(lambda y:jv(1,2*np.sqrt(x[k]*y/c))*np.sqrt(x[k]/y)
                                       *hrH[j](y)*np.exp(-y**2/2),0,np.inf)[0]

p=[]
beta=np.zeros(2*J)
for z in zs:
 sol=root(eqs,beta,args=(z),tol=1e-12)
 beta=sol.x
 betac=beta[:J]+1j*beta[J:]
 f_real=lambda y: np.real(np.exp(-1j*y/np.sqrt(c)-y**2/2)*np.sum([betac[j]
                      *hrH[j](y) for j in range(J)]))
 f_imag=lambda y: np.imag(np.exp(-1j*y/np.sqrt(c)-y**2/2)*np.sum([betac[j]
                      *hrH[j](y) for j in range(J)]))
 p.append(np.real(quad(f_real,0,np.inf)[0]
                     +1j*quad(f_imag,0,np.inf)[0])/(np.pi*sqrtc))

plt.plot(c+1+zs*sqrtc,p)
plt.xlim(0)
plt.ylim(0)
plt.show()
\end{verbatim}


\end{document}